\documentclass[aps,pra,showpacs,twocolumn]{revtex4}
\pdfoutput=1
\usepackage[ansinew]{inputenc}
\usepackage{amsmath}
\usepackage{graphicx}
\usepackage{amsmath, amssymb}
\renewcommand{\vr}{\ensuremath{\mathbf{r}}}
\newcommand{\vp}{\ensuremath{\mathbf{p}}}
\newcommand{\chicm}{\ensuremath{\chi_\mathrm{cm}}}
\newcommand{\Id}{\mathbb{I}}
\newcommand{\aosc}{\ensuremath{a_\text{osc}}}
\newcommand{\matom}{\ensuremath{m_\text{atm}}}
\renewcommand{\Re}{\mathrm{Re}}
\renewcommand{\Im}{\mathrm{Im}}
\begin{document}
\title{Pair correlated atoms with a twist}
\author{Uffe V. Poulsen and Klaus M\o lmer}
\affiliation{Lundbeck Foundation Theoretical 
  Center for Quantum System Research\\
  Department of Physics and Astronomy\\
  University of
  Aarhus\\
  DK 8000 Aarhus C, Denmark}
\date{\today}
\begin{abstract}
  We present an analysis of the quantum state resulting from the
  dissociation of diatomic molecules prepared in a condensate vortex
  state. The many-body state preserves the rotational symmetry of the
  system in quantum correlated states by having two equally populated
  components with angular momentum adding to unity. A simple two-mode
  analysis and a full quantum field analysis is presented for the case
  of non-interacting atoms and weak depletion of the molecular
  condensate.
\end{abstract}
  \maketitle
\section{Introduction}
Since the 1995 experiments with the first production of atomic
Bose-Einstein condensates, degenerate quantum gasses have constituted
a very active field of research. A wide variety of means exists for
detection and control of the properties of these systems, and the
early works have been followed by progress on degenerate fermionic
systems, on mixtures of different species, and on conversion between
atomic and molecular quantum gasses.

The coherence properties of the degenerate quantum states have been
verified implicitly by measurements of the response properties of the
systems and explicitly by the observation of robust
interferences~\cite{andrews97:_obser_inter_between_two_bose_conden}
and topological
structures~\cite{madison00:_vortex_format_stirr_bose_einst_conden}.
Recently, a vortex lattice in a system of fermionic atoms was shown to
extend over the BCS-BEC crossover towards formation of bosonic
diatomic molecules on the molecular side of a field controlled
Feschbach resonance~\cite{zwierlein05:_vortices_fermi}.

Many properties of condensates are well described by mean field
theories and the Gross-Pitaevskii equation, but in some cases mean
field theories may completely fail and totally forbid processes, which
occur peacefully according to a full quantum state analysis. Important
examples of such processes are the period-doubling observed in a
shaken lattice~\cite{gemelke05:_param_amp_matter_waves} and the
break-up of a moving condensate in a lattice due to four wave
mixing~\cite{campbell06:_param_amp_atom_pairs}. Both processes are
driven by collisions of pairs of condensate atoms which emerge in two
new momentum states. Within a mean field theory, these processes, like
the equivalent parametric amplification process in optical
down-conversion and four wave mixing will only get initiated if a
non-vanishing mean field with the final state character is seeded to
the solution of the Gross-Pitaevskii
equation~\cite{hilligsoe05:_four_wave_mixing}. From a formal
perspective, the fact that only the sum of the phases of the two final
state components and not the individual phases are locked to the
initial state wave functions causes the fields to remain in the
vacuum state in order not to break the phase symmetry. If the physical
problem is simple, e.g., in the few-mode quantum optical problems, one
may alternatively have recourse to a full quantum many-body
theoretical analysis. In the context of quantum gasses, we have
previously~\cite{poulsen01:_quant_states_bose_einst} studied the
"degenerate down-conversion" of a molecular condensate by dissociation
to a single trapped atomic condensate in such a full quantum treatment
and shown that despite the absence of an atomic condensate phase, the
atomic component does indeed accumulate in a single preferred quantum
state. Dissociation into two atomic beams have smilarly been proposed
as a means for generation of Einstein-Podolsky-Rosen
correlations~\cite{kheruntsyan05:_epr_dis_mol_bec}.

Note that in four-wave mixing processes a mean-field solution not only
breaks the phase symmetry of the field, it also breaks the
translational invariance of the problem as dramatically observed as
period doubling in~\cite{gemelke05:_param_amp_matter_waves}, and
likewise the different populated momentum components
in~\cite{campbell06:_param_amp_atom_pairs} will have spatial
interference patterns with periods exceeding the one of the lattice
potential used in the experiments. In this paper, we consider a
process where the symmetry breaking is similarly spectacular, namely
the down conversion by photo dissociation of a molecular condensates
which is prepared in a single vortex state. In a cylindrically shaped
trap, a vortex state is a topologically stable state of a quantum
gas with a vanishing density along the cylinder axis and a phase which
changes by $2\pi$ as one follows a closed loop around the axis. The
rotational symmetry around the $z$ axis is a fundamental property of
the many-body Hamiltonian and both mean field solutions and more
elaborate theories must respect this symmetry.  Microscopically, the
vortex solution is consistent with a Hartree product state description
with a product of single particle $M=1$ eigenstates of the azimuthal
angular momentum.  In this state, every molecule has a unit angular
momentum around the condensate axis, which cannot be transferred to a
pair of atoms populating the same ($m=1/2$ ?) quantum state.
Conversely, any mean field solution for the atomic system will
necessarily correspond to an integer angular momentum per atom, and
would hence imply an even angular momentum of every pair, i.e, of the
molecule.

The molecules must dissociate to at least two different atomic states
with azimuthal quantum numbers which add to the molecular value,
$m+m'=M$. The easiest situation to deal with is one where only two
states with angular momentum $m=0$ and $m'=1$ get populated. We shall
address this case in Sec. II. In particular we shall discuss to which
extent a strongly number correlated quantum state of this kind state
is distinguishable from a state which macroscopically populates an
even weight superposition of the same two single particle states. The
restriction to only two finally occupied states should be justified by
a more elaborate treatment of the many-body Hamiltonian, and in Sec.
III, we model the actual process in which the atoms are not put
directly into two states extending over vast and partly
non-overlapping regions of the atomic single particle quantum states,
but where the molecular dissociation enforces a localized common
origin of the matter waves. The problem is solved exactly by a
Bogoliubov tranformation, and we identify different regimes and the
validity of the few-mode Ansatz.

\section{Two-mode analysis}
\label{sec:two-mode_analysis}
For simplicity we ignore interaction between atoms, between molecules,
and between atoms and molecules. We also make the assumption that only
one quantum state for the molecules and two quantum states for the
atoms are relevant. The molecule state should be the macroscopically
occupied state of the molecular condensate. The two atomic states are
assumed to be selected by a resonance condition: They could for
example be the two lowest 2D single particle eigenfunctions in a
harmonic trap [cf.
Eq.~(\ref{eq:def_phi_nmk}) below], $\Phi_{00}(x,y)\propto
\exp(-\rho^2/2\aosc^2),\Phi_{01}(x,y)\propto
(x+iy)\exp(-\rho^2/2\aosc^2)$, where $\rho^2=x^2+y^2$, and where the
oscillator length $\aosc=\sqrt{\hbar/\matom \omega}$ is defined in
terms of the atomic mass $\matom$ and the trap frequency $\omega$.
The lowest state $\Phi_{00}$ has angular momentum $m=0$, while
$\Phi_{01}$ has $m'=1$ so that they fulfill the condition for angular
momentum conservation when molecules with $M=1$ are dissociated,
$m+m'=M$.

If we can drive the system exactly at resonance, i.e., either the
atomic and molecular states are degenerate, or in a laser induced
dissociation process, the field frequencies involved exactly fulfill
the Bohr frequency condition for quantum transitions, our model
Hamiltonian becomes
\begin{equation}\label{Hsim}
\hat{H}
=
\beta\, \hat{c} \hat{a}_0^\dagger \hat{a}_1^\dagger
+\beta^*\, \hat{c}^\dagger \hat{a}_0\hat{a}_1
.
\end{equation}
Here $\beta$ quantifies the strength of the microscopic coupling, the
operator $\hat{c}$($\hat{c}^\dagger$) annihilates (creates) a
molecule, and the operators
$\hat{a}_0,\hat{a}_1$($\hat{a}_0^\dagger,\hat{a}_1^\dagger$)
annihilate (create) atoms in the two atomic states. This Hamiltonian
is well known in quantum optics where it describes the non-degenerate
optical parametric oscillator. In general it leads to a complicated
entanglement between the pump beam ($\hat{c}$) and the signal and idler
beams $\hat{a}_0,\hat{a}_1$ which can of course be made subject to
detailed investigation~\cite{dechoum04:_nondegen_para_oscillator}, but
in the limit of a strong pump, the depletion can often be ignored and
the pump operators can be replaced by c-numbers. We will assume here
that the molecular condensate is sufficiently large that depletion can
be neglected, and hence we shall consider the simpler Hamiltonian
\begin{equation}\label{Hsimple}
\hat{H}
=
\chi \hat{a}_0^\dagger \hat{a}_1^\dagger
+\chi^* \hat{a}_0\hat{a}_1 
\end{equation}
The coupling strength $\chi$ now includes the c-number describing the
molecular field.  Note that the quadratic Hamiltonian (\ref{Hsimple})
does not conserve the number of atoms, and the final state appears to
be a coherent superposition of states with different numbers of atoms.
This is, however, only because we omit the meticulous writing of the
associated molecular components of the states, which precisely account
for the conservation of atom numbers.  As we shall only be interested
in number conserving observables such as the atomic density
distribution, we shall make no errors in applying the symmetry
breaking Hamiltonian (\ref{Hsimple}).

Starting in a state with no atoms, the atomic vacuum state, in both
modes, the time evolution leads to the production of the two-mode
state
\begin{equation} \label{squeezed}
|\Psi\rangle
= \sqrt{1-|s|^2}
\sum_n  s^n
|n,n\rangle,
\end{equation}
where
\begin{equation}
  \label{eq:def_s}
  s = -i\frac{\chi}{|\chi|} \tanh(|\chi| t/\hbar).
\end{equation}
As discussed in detail in the Introduction, this state does not follow
from a mean field analysis and, indeed, the mean values of the field
operators $\hat{a}_i$ vanish exactly, whereas the state has a mean atom
number of
\begin{equation}
  \label{eq:N_of_t}
  \langle \hat{N}_i \rangle 
  =
  \langle \hat{a}_i^\dagger \hat{a}_i \rangle
  =
  |s|^2/(1-|s|^2)
  =
  \sinh^2(|\chi|t/\hbar)
\end{equation}
in each mode. The atom number distributions are exponential (thermal)
and thus the fluctuations in the atom number are large,
$\mathrm{Var}( \hat{N}_i)=\langle \hat{N}_i\rangle^2+\langle
\hat{N}_i\rangle$.

A number of papers have discussed observational difference between
systems populating several single particle states macroscopically and
a coherent superposition of the same states. It has been
argued~\cite{javanainen96:_quant_phase_bose_einst_conden,castin97:_relat_bose_einst,moelmer97:_opt_coh_fiction,cirac96:_contin_bose}
that such differences are small or insignificant for systems with many
atoms. The general argument was supplemented by simulations
of actual detection records, where the back action due to local
measurements on the system turned out precisely to establish a
definite relative phase of the two components.

The restriction to only two modes implies that the one-body density
matrix is fully characterized by the 2x2 matrix with
$\rho_{ij}=\langle a^\dagger_i a_j\rangle$. In particular the spatial
density is given by
\begin{equation}
\label{dens}
\begin{split}
n(x,y)
&=
\rho_{00}|\Phi_{00}(x,y)|^2
+\rho_{11}|\Phi_{01}(x,y)|^2
\\
&\phantom{=}+2\Re[\rho_{01}\Phi_{00}(x,y)\Phi_{01}^*(x,y)].
\end{split}
\end{equation}
Registration of a particle at position $(x_d,y_d)$, chosen according
to this probability distribution, causes the application of the
annihilation operator
$\Phi_{00}(x_d,y_d)\hat{a}_0+\Phi_{01}(x_d,y_d)\hat{a}_1$ on the state
vector expanded in the two mode number components as in (3), followed
by a renormalization. We are thus able to compute the up-dated 2x2
density matrix and by repeating these steps to simulate the subsequent
detection of a number $k$ of atoms. The present problem differs in two
way from the simulations reported in
Refs.~\cite{javanainen96:_quant_phase_bose_einst_conden,moelmer97:_opt_coh_fiction}:
the two modes populated are not plane waves but they have different
spatial dependencies, implying that some atomic detection events can
be ascribed solely to one component (e.g., only the $m=0$ component
contributes to the atomic density on the vortex axis), and the inital
state does not have a well defined number of atoms, but much larger
than Poissonian fluctuations, and hence the rigid mathematical
analysis of the emergence of a relative phase in
Ref.~\cite{castin97:_relat_bose_einst} does not apply. We have carried
out simulations, and despite these two differences, the 2x2 density
matrix indeed approaches a pure state projector and we obtain
detection patterns that are compatible with the density of a single
coherent quantum state.

\section{Multi-mode analysis}
\label{sec:many_mode_analysis}
Let us now analyze the problem without making the two-mode
simplification. Retaining the approximation of replacing molecule
operators with c-numbers, a full many-mode version of
Eq.(\ref{Hsimple}) reads
\begin{equation}
  \label{eq:H_many_modes}
  \begin{split}
    H =& \int\!d^3\!r\; \hat\Psi^\dagger(\vr)h(\vr)\hat\Psi(\vr)
    \\
    &+ \left\{ \int\!d^3\!rd^3\!r'\;
      \chi(\vr,\vr')\hat\Psi^\dagger(\vr)\hat\Psi^\dagger(\vr') +
      \text{h.c.}  \right\} .
  \end{split}
\end{equation}
The single particle part consists of kinetic energy and
external trapping potential
\begin{equation}
  \label{eq:def_h}
  h(\vr)=\frac{\vp^2}{2\matom} + V_\mathrm{ex}(\vr)-\Delta_\text{bare}
  .
\end{equation}
The energy offset $\Delta_\text{bare}$ is half of the two-atom
detuning: the energy of a trapped molecule in the vortex state is the
energy of two free atoms plus $2\Delta_\text{bare}$.

Typical traps are well approximated by a harmonic potential
$V_\mathrm{ex}(\vr)=\matom(\omega_x^2 x^2+\omega_y^2 y^2 + \omega_z^2
z^2)/2$. We will consider situations with axial symmetry and we
therefore let $\omega_x=\omega_y=\omega$. For the purpose of
discussing the dissociation of a molecular condensate with a vortex
along $z$, it is useful to make one of two simplifying assumptions
about the dynamics along $z$:
\begin{itemize}
\item A quasi cylindrical situation with $\omega_z \ll
  \omega_\perp$ such that the $z$ component of the momentum,
  $p_z$, becomes approximately conserved.
\item A quasi two-dimensional situation where $\omega_z \gg
  \omega_\perp$ so that the dynamics along $z$ is effectively
  frozen to take place in a single quantum state.
\end{itemize}
We focus on the second case in the following. The modifications needed
to treat the first case are briefly discussed in the Appendix.

The Hamiltonian~(\ref{eq:H_many_modes}) can describe many processes
where pairs of particles are created: one must choose the
correct coupling kernel $\chi(\vr,\vr')$ for the problem under
consideration. When dealing with dissociation of molecules,
$\chi(\vr,\vr')$ depends on the wavefunction of individual molecules
(both relative motion and center-of-mass), and on the possible spatial
variation of the external fields mediating the dissociation. We will
assume the relative motion molecular wavefunction to be very well
localized compared to other length scales in the problem. This means
that the dependence of $\chi(\vr,\vr')$ on the separation $\vr-\vr'$
can be approximated by a delta-function. We further assume that the
center-of-mass molecular wavefunction and the relevant external fields
are rotationally symmetric around the $z$-axis so that
\begin{equation}
  \label{eq:chi_delta}
  \chi(\vr,\vr')
  =
  \delta(\vr-\vr')\chicm(\rho,z)e^{iM\phi}
  ,
\end{equation}
in usual cylindrical coordinates where $x=\rho\cos\phi$ and
$y=\rho\sin\phi$. In the simplest case the dissociating external
fields are spatially uniform over the extend of the molecular
condensate. Then $M$ is the charge of the molecular vortex state and
the remaining spatial variation of $\chicm(\rho,z)$ is simply given by
the norm of the c-number field that describes the molecular condensate
(squareroot of the density)~\footnote{In principle, the dissociating
  fields could also contribute to $M$. Transfer of (orbital) angular
  momentum in Bragg scattering of BECs has been demonstrated in
  Ref.~\cite{andersen:_quant_rot}.}. To specify the overall strength
of the coupling a microscopic model of the dissociation must be
decided upon.  In Ref.~\cite{heinzen00:_super}, a two-photon Raman
process between a bound molecular state and ``free'' (except for
external trapping) atoms is considered. The dissociation is via an
excited molecular state, which should ideally be only negligibly
populated since spontaneous decay from it will be an unwanted loss
mechanism. The particular example in Ref.~\cite{heinzen00:_super}
shows that a peak value of our $\chi$ comparable to the trap level
spacing is fully realistic. This is the regime we focus on below.

We emphasize that this physical modelling does not favor any
particular pair of atomic modes, and the
Hamiltonian~(\ref{eq:H_many_modes}) cannot be brought on the
form~(\ref{Hsimple}).  Atoms are coherently prepared from the entire
region populated by the molecules, but pairs are initially much more
tightly located both radially and azimuthally than the single mode
functions suggested in the previous section. In particular, the atoms
are not restricted to low values of the angular momentum quantum
number. The temporal evolution of the system due to the kinetic energy
and external potential, and the role of energy conservation, however,
favors the population of only few modes as we shall see below.

\subsection{Frozen $z$ dynamics}
\label{sec:frozen_z} In the case of tight $z$ confinement for the
atoms, we get a simple description.  Let the only accessible atomic
$z$ mode be $\phi_z$, $\int dz |\phi_z|^2=1$. We can then expand the
atomic field operators on the discrete set of mode operators
\begin{equation}
  \label{eq:Psi_one_z}
  \hat\Psi(\vr)
  =
  \sum_{nm} \hat{a}_{nm}
  \times
  \Phi_{nm}(\rho)
  \times
  \sqrt{\frac{1}{2\pi}} e^{im\phi}
  \times
  \phi_z(z)
\end{equation}
where the radial modefunction $\Phi_{nm}$ are
\begin{equation}
  \label{eq:def_phi_nmk}
  \begin{split}
    \Phi_{nm}&(\rho) = \sqrt{\frac{n!}{2\left( n+|m|
        \right)!\aosc^{2}}}
    \\
    &\times
    \exp\left(-\frac{1}{2}\frac{\rho^2}{\aosc^2}\right)
    \left(\frac{\rho}{\aosc}\right)^{|m|}
    L^{|m|}_n\left(\frac{\rho^2}{\aosc^2}\right) ,
  \end{split}
\end{equation}
with $L^m_n$ the $m$'te associated Laguerre polynomum of order $n$ and
$\aosc=\sqrt{\hbar/\matom\omega}$ is the oscillator length.  The
commutation relations of the $\hat{a}_{nm}$ are
\begin{equation}
  \label{eq:froz_rad_commu}
  \begin{split}
    \Bigl[\hat{a}_{nm},\hat{a}_{n'm'}\Bigr]
    &=
    0
    \\
    \Bigl[\hat{a}_{nm},\hat{a}^\dagger_{n'm'}\Bigr]
    &=
    \delta_{nn'}\delta_{mm'}
    .
  \end{split}
\end{equation}
The Hamiltonian becomes
\begin{multline}
  \label{eq:H_frozen}
  \hat{H} = \sum_{nm} E_{nm} \hat{a}^\dagger_{nm}\hat{a}_{nm}
  \\
  + \sum_{m} \sum_{nn'} \left\{ K_{nn'm}
    \hat{a}^\dagger_{nm}\hat{a}^\dagger_{n'(M-m)} +\text{h.c.}
  \right\} ,
\end{multline}
where $K_{nn'm}$ is defined via\begin{equation}
  \label{eq:def_K_nnm}
  \begin{split}
    &K_{nn'm}
    =
    \\
    &\int\! d\!z d\!\rho \rho  \;
    \chi_\mathrm{cm}(\rho,z) \;
    |\phi(z)|^2 \;
    \Phi^*_{nm}(\rho) \Phi_{n'M-m}(\rho)
    .
  \end{split}
\end{equation}
and $E_{nm}$
is given by
\begin{equation}
  \label{eq:E_nm}
  E_{nm}=\left(2n+|m|+1\right)\hbar\omega-\Delta
  .
\end{equation}
Here $\Delta$ is the effective detuning, adjusted for the energy
associated with $\phi_z$:
\begin{equation}
  \label{eq:def_Delta}
  \Delta=\Delta_\text{bare}-\int \!dz\;
  \phi_z^*\left(\frac{p_z^2}{2m}+\frac{1}{2}m\omega_z^2z^2\right)\phi_z
  .
\end{equation}

\section{Bogoliubov diagonalization}
\label{sec:bogo_diag}
The Hamiltonian (\ref{eq:H_frozen}) is a quadratic form of creation
and annihilation operators and as such it can in general be decoupled
to a collection of independent harmonic oscillators by a
\emph{Bogoliubov transformation}. Please note that this transformation
is usually applied in connection with the \emph{Bogoliubov
  approximation} in theoretical studies of Bose-Einstein condensed
systems. The Bogoliubov approximation provides a quadratic Hamiltonian
by appeal to the macroscopic population of the condensate mode. Here
the quadratic form of the Hamiltonian is rather a consequence of (i)
that two atoms are created in each fundamental dissociation process
and (ii) that we ignore the dynamics (in particular the depletion) of
the molecules and describe them by a stationary c-number field. One
could in principle extend our approach to also treat the interaction
among the created atoms in a Bogoliubov approximation, but since we
focus on situations with initially \emph{no} atoms present this would
neither be necessary nor indeed well justified. For an example of
\emph{time-dependent} quadratic Hamiltonians, see the work by Zi\'{n}
\textit{et al}.\ on colliding
condensates~\cite{zin05:_q_multimode_elast_scat,zin06:_elast_scat_bec}.

\subsection{Two-mode Bogoliubov}
\label{sec:two_mode_bogo} 
Let us first look at the simple case of two modes like in
Sec.~\ref{sec:two-mode_analysis}. In addition to the pair creation and
annihilation terms of Eq.~(\ref{Hsimple}) we include mode energies
$\hbar\omega_0$ and $\hbar\omega_1$:
\begin{equation}
  \label{eq:Hsimple_delta}
  \begin{split}
    \hat{H}
    &=
    \hbar\omega_0 \left( \hat{a}_0^\dagger \hat{a}_0+\frac{1}{2}\right)
    + \hbar\omega_1 \left(\hat{a}_1^\dagger \hat{a}_1+\frac{1}{2}\right)
    \\
    &\phantom{=}
    +\chi \hat{a}_0^\dagger \hat{a}_1^\dagger+\chi^* \hat{a}_0 \hat{a}_1
    .
  \end{split}
\end{equation}
A useful and compact notation is
\begin{equation}
  \label{eq:H_simple_compact}
  \hat{H}
  =
  \frac{1}{2}\hat{A}^\dagger \mathrm{h} \hat{A}
\end{equation}
with
\begin{equation}
  \label{eq:def_A}
  \hat{A}
  =
  \begin{bmatrix}
    \hat{a}_0 \\
    \hat{a}_1 \\
    \hat{a}_0^\dagger \\
    \hat{a}_1^\dagger
  \end{bmatrix}
  ,
  \quad
  A^\dagger
  =
  \begin{bmatrix}
    \hat{a}_0^\dagger &
    \hat{a}_1^\dagger &
    \hat{a}_0 &
    \hat{a}_1
  \end{bmatrix}
  ,
\end{equation}
and
\begin{equation}
  \label{eq:def_h_rm}
  \mathrm{h}
  =
  \begin{bmatrix}
    \hbar\omega_0 & 0 & 0 & \chi \\
    0 & \hbar\omega_1 & \chi & 0 \\
    0 & \chi^* & \hbar\omega_0 &0 \\
    \chi^* & 0 & 0 & \hbar\omega_1
  \end{bmatrix}
  .
\end{equation}
We now seek to simplify $\hat{H}$ by a Bogoliubov transformation,
i.e.\ we define new creation and annihilation operators implicitly by
\begin{equation}
  \label{eq:def_bogo}
   \begin{bmatrix}
    \hat{a}_0 \\
    a_1 \\
    \hat{a}_0^\dagger \\
    a_1^\dagger
  \end{bmatrix}
  =
  \begin{bmatrix}
    U & V^* \\
    V & U^*
  \end{bmatrix}
  \begin{bmatrix}
    \hat{b}_0 \\
    \hat{b}_1 \\
    \hat{b}_0^\dagger \\
    \hat{b}_1^\dagger
  \end{bmatrix}
\end{equation}
where the 2$\times$2 matrices $U$ and $V$ should fulfill
\begin{equation}
  \label{eq:props_UV}
  \begin{split}
  U^\dagger U-V^\dagger V&=\Id_2
  \\
  U^\mathrm{T}V-V^\mathrm{T}U&=0
\end{split}
\end{equation}
in order for $\hat{b}_0$ and $\hat{b}_1$ to have the commutation
relations for independent bosonic creation and annihilation operators
[see Eq.~(\ref{eq:froz_rad_commu})]. Note that in general the
Bogoliubov transform is \emph{not} simply a choice of different
spatial modes.

\subsubsection{Detuning dominated case $2|\chi|<|\hbar\omega_0+\hbar\omega_1|$}
\label{sec:detuned}
Depending on the strength of the coupling $|\chi|$ relative to the
two-atom detuning $|\hbar\omega_0+\hbar\omega_1|$ the Hamiltonian can
be written in one of two standard forms. If
$2|\chi|<|\hbar\omega_0+\hbar\omega_1|$, $\hat{H}$ can be written as a
sum of two independent oscillators
\begin{equation}
  \label{eq:H_real}
  \hat{H}
  =
  \lambda_0 \left(
   \hat{b}_0^\dagger \hat{b}_0 + \frac{1}{2}
  \right)
  +\lambda_1 \left(
    \hat{b}_1^\dagger \hat{b}_1 + \frac{1}{2}
  \right)
  ,
\end{equation}
with
\begin{equation}
  \label{eq:lambdas_real}
  \begin{split}
  \lambda_{0,1}
  =&
  \pm\frac{1}{2}(\hbar\omega_0-\hbar\omega_1)
  \\
  &  +
  \frac{1}{2}(\hbar\omega_0+\hbar\omega_1)
  \sqrt{1-\frac{4|\chi|^2}{|\hbar\omega_0+\hbar\omega_1|^2}}
\end{split}
\end{equation}
The Bogoliubov transformation is given by:
\begin{equation}
  \label{eq:U_decoupled}
  U
  =
  \begin{bmatrix}
    \cosh r & 0 \\
    0 & \cosh r
  \end{bmatrix}
\end{equation}
and
\begin{equation}
  \label{eq:V_decoupled}
  V
  =
  \begin{bmatrix}
    0 & \mp \frac{\chi^*}{|\chi|} \sinh r \\
    \mp\frac{\chi^*}{|\chi|}\sinh r & 0
  \end{bmatrix}
  ,
\end{equation}
where the upper (lower) sign should be chosen for
$\hbar\omega_0+\hbar\omega_1$ positive (negative). The squeezing
parameter $r$ is defined through
\begin{equation}
  \label{eq:r_decoupled}
  \tanh r =
  \frac{2|\chi|}{|\hbar\omega_0+\hbar\omega_1|
    +\sqrt{|\hbar\omega_0+\hbar\omega_1|^2-4|\chi|^2}}
\end{equation}
We see that as $2|\chi|$ approaches $|\hbar\omega_0+\hbar\omega_1|$, the
coefficients in the transformation diverge, i.e.\ the decoupled modes
become infinitely squeezed.

Because of the diagonalized form of the Hamiltonian (\ref{eq:H_real}),
the time evolution will be independent for the two Bogoliubov modes:
Each will simply behave as a harmonic oscillator.  This means that the
dynamics of all quantities will be oscillatory with two fundamental
frequencies given by $\lambda_{0,1}/\hbar$.

Note, that even if both mode energies $\hbar\omega_{0,1}$ are positive so
that a pair of atoms actually has a higher energy than a molecule, one
of the eigenvalues $\lambda_i$ can become negative. This happens if
$|\chi|^2> \hbar\omega_0\hbar\omega_1$ and in that case the system will be
thermodynamically unstable as it is energetically favourable to create
the corresponding kind of quasi-particles.

\subsubsection{Coupling dominated case $2|\chi|>|\hbar\omega_0+\hbar\omega_1|$}
\label{sec:coupled} 
If $2|\chi|>|\hbar\omega_0+\hbar\omega_1|$, the simplest form attainable by
Bogoliubov transformations involves pair-creation (with real positive
coefficient) into two symmetrically detuned modes:
\begin{equation}
  \label{eq:H_complex}
  \hat{H}
  =
  \Re[\lambda] \left( \hat{b}_1^\dagger \hat{b}_1 
    - \hat{b}_0^\dagger \hat{b}_0 \right)
  +\Im[\lambda]\left( \hat{b}_0^\dagger \hat{b}_1^\dagger 
    + \hat{b}_0 \hat{b}_1 \right)
\end{equation}
with
\begin{equation}
  \label{eq:lambdas_complex}
  \begin{split}
    \lambda 
    =&
    \frac{1}{2}(\hbar\omega_1-\hbar\omega_0)
    \\
    &+ \frac{i}{2}(\hbar\omega_0+\hbar\omega_1)
    \sqrt{\frac{4|\chi|^2}{|\hbar\omega_0+\hbar\omega_1|^²}-1}
    .
  \end{split}
\end{equation}
The transformation is still of the form given in
Eqs.~(\ref{eq:U_decoupled}) and (\ref{eq:V_decoupled}), but now the
squeezing parameter $r$ should be found from:
\begin{equation}
  \label{eq:r_pairs}
  \tanh r =
  \frac{|\hbar\omega_0+\hbar\omega_1|}
  {2|\chi|+\sqrt{4|\chi|^2-|\hbar\omega_0+\hbar\omega_1|^2}}
\end{equation}
We see that as before, $r$ diverges when $2|\chi|$ approaches
$|\hbar\omega_0+\hbar\omega_1|$. At exact two-\emph{atom} resonance, i.e.
$\hbar\omega_0+\hbar\omega_1=0$, the Bogoliubov operators are simply the
original $\hat{a}_0$ and $\hat{a}_1$.

The transformed Hamiltonian (\ref{eq:H_complex}) consist of two
commuting terms and the first one even conserves the number of
quasi-particles $\hat{N}_b=\hat{b}^\dagger_0 \hat{b}_0 +
\hat{b}^\dagger_1 \hat{b}_1$. The second term leads to unbound
creation of quasi-particle pairs, in fact $\langle \hat{N}_b\rangle$
will grow as $\sinh^2 (\Im[\lambda] t)$. This production of
quasi-particles translates to a similarly unbounded growth in the
number of atoms (until depletion of the molecular condensate renders
our simple Hamiltonian invalid), and because of the Bogoliubov
tranformation, the atom numbers will show oscillations around their
general growth.

\subsection{Multi-mode Bogoliubov}
\label{sec:many_modes_bogo}
The generalization to many modes is relatively straightforward. By
Bogoliubov transformations of the original set of creation and
annihilation operators, $\hat{H}$ can be written as a sum of a number of
independent oscillators and a number of pairs of symmetrically detuned
modes into which quasi-particles are created:
\begin{equation}
  \label{eq:H_many_modes_diag}
  \begin{split} 
    \hat{H} 
    = &
    \sum_{j} \lambda_j 
    \left( 
      \hat{b}j^\dagger \hat{b}_j + \frac{1}{2} 
    \right)
    \\
    &+ 
    \sum_{p} \Re[\lambda_p] 
    \left( 
      \hat{b}_{p2}^\dagger \hat{b}_{p2}
      - \hat{b}_{p1}^\dagger \hat{b}_{p1} 
    \right)
    \\
    &\phantom{+\sum_{p}}
    +\Im[\lambda_p]
    \left( 
      \hat{b}_{p1}^\dagger \hat{b}_{p2}^\dagger 
      + \hat{b}_{p1} \hat{b}_{p2}
    \right) .
  \end{split}
\end{equation}
The dynamics in the decoupled modes is oscillatory while ever more
quasi-particles will be created in the paired modes. Note that it is
perfectly possible that all modes can be decoupled so that the unitary
dynamics is purely oscillatory.

\subsection{Consequences of the azimuthal symmetry}
\label{sec:symmetries}

We consider a situation with axial symmetry: The single particle part
of the Hamiltonian, $h(\vr)$ of Eq.~(\ref{eq:def_h}), is invariant
under rotations around the $z$-axis. This leads to the usual
block-diagonal form of the single particle Hamiltonian with respect to the
quantum number $m$, the eigenvalue of the $z$-component of the angular
momentum operator. At the same time, the molecular field [$\chi$ of
Eq.~(\ref{eq:chi_delta})] changes simply by a phase factor
$\exp(iM\phi)$ when the system is rotated by an angle $\phi$. This
leads to an exact selection rule on the pairs of modes that are
populated in the dissociation process, namely $m+m'=M$. For odd $M$
this imposes an ``off-diagonal'' form on the pair creation, as the
atoms created by a dissociation process must necessarily end up with
different integer quantum numbers, $m$ and $M-m$, whereas for even $M$
the two atoms can both end up in an $m=M/2$ mode.

In summary, each $m$ subspace is coupled only to itself by the single
particle part of the Hamiltonian and the dissociation can either
couple it exclusively to itself or exclusively to one other $m$
subspace. In this way the full problem splits into a series of smaller
problems consisting of either one or two $m$-subspaces.

Now consider the problem of two coupled subspaces. As pairs are always
created with an atom in each subspace, there is an obvious
conservation of the \emph{difference} in the total number of atoms in
the two subspaces. This can also be phrased as the invariance of the
evolution under shifts in the \emph{relative} energy of the two
subspaces as long as their total energy is maintained. Among other
consequences, this implies that there are no first-order coherences
between the two subspaces and that the one-body density operator is
correspondingly block-diagonal. It is, in fact, possible to show an
even stronger, dynamical symmetry in the evolution: If the two
subspaces are initially unpopulated, they will at all times remain
unitarily equivalent. In particular, their one-body density matrices
will have identical spectra of
eigenvalues~\cite{poulsen:_in_preparation}.

\section{Results}
\label{sec:results}
\begin{figure}[htbp]
  \centering
  \includegraphics{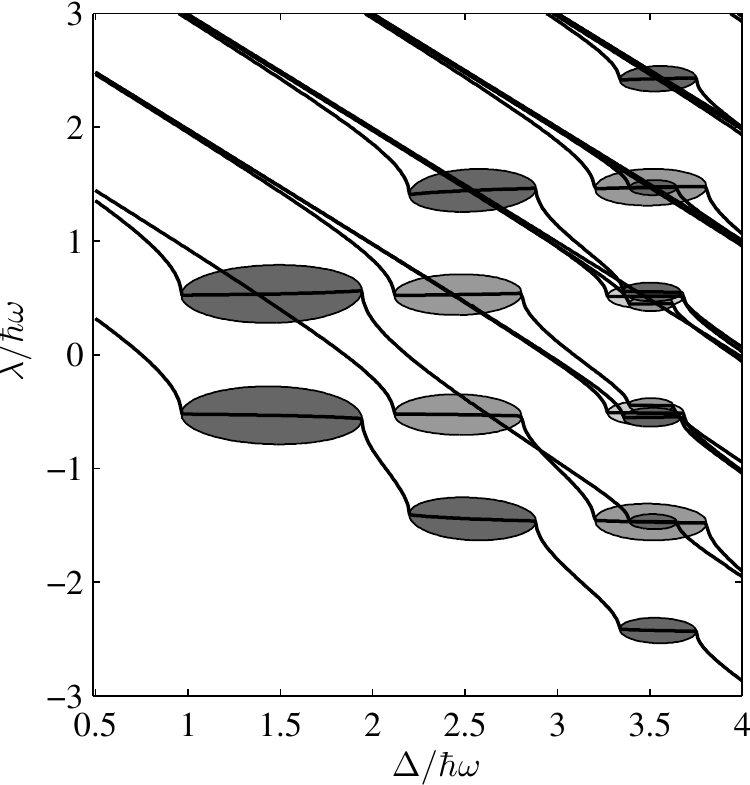}
  \caption{Bogoliubov eigenvalues as functions of the detuning
    $\Delta$ for the Hamiltonian~(\ref{eq:H_frozen}) with a coupling
    field (\ref{eq:chi_delta}) with $M=1$. For concreteness, we assume
    a strength and a radial dependence of the coupling such that
    $\int\!  d\!z\chi_\mathrm{cm}|\phi_z|^2 = 2 \hbar\omega \rho
    \exp(-\rho^2/\aosc^2)/\aosc$. The lines show $\Re(\lambda)$, which
    are the mode energies while, in the coupling dominated regime, the
    shaded areas around them have a height of $\Im(\lambda)$ to
    indicate the coupling between pairs of modes.  The darkness of the
    shading indicates the $m$-values in the pairs of coupled modes:
    the darkest shading is for $m=0,m'=1$ pairs, $m=-1,m'=2$ pairs are
    a tone brighter, etc. Note that members of coupled pairs lie
    symmetrically around $\lambda=0$.}
  \label{fig:lambdas}
\end{figure}

As discussed above, once the Hamiltonian has been brought to a
standard form by a Bogoliubov transformation, a number of properties
of the system are immediately clear. In particular, the oscillatory or
unbounded behaviour depends on whether all modes can be decoupled. In
Fig.~\ref{fig:lambdas} we show the real and imaginary parts of the
$\lambda_j$'s as a function of the detuning $\Delta$.
When $\Delta$ is increased, all atomic modes move linearly down in
energy according to the single atom Hamiltonian (7). As a pair of
modes with $m+m'=M=1$ gets close to the two-atom resonance, i.e. to
having equal and opposite energies, the coupling has the effect of
bending the quasi-particle energy levels further downwards. The modes
become increasingly squeezed but remain uncoupled -- they are in the
detuning dominated regime of Sec.~\ref{sec:detuned}. At the point of
perfect two-quasi-particle resonance, it is no longer possible to
decouple the two modes and the system moves into a regime with
quasi-particle pair production, the coupling dominated case of
Sec.~\ref{sec:coupled}.  When a pair of quasi-particle modes become
unstable, it is signalled by the corresponding $\lambda_j$'s attaining
a finite imaginary part. In Fig.~\ref{fig:lambdas} this is plotted as
a shaded region around the corresponding energy curves. The height of
the shaded region signifies the magnitude of $\Im[\lambda]$. As
$\Delta$ is further increased, the pair production strength also
increases. It has its maximum approximately at the point of
two-\emph{atom} resonance, i.e., when the two \emph{bare} atomic modes
becomes resonant with the molecule energy.

As an example, let us follow the lowermost curve through the figure.
At the left, at low values of $\Delta$, the curve corresponds to the
atomic harmonic oscillator ground state in the trap with $n=0$ and
$m=0$. When $\Delta$ gets close to 1, pair production together with
the ($n=0,m=1$) state (second lowest curve) becomes resonant. Note
that the ($n=0,m=1$) state is degenerate with the ($n=0,m=-1$) for
very low $\Delta$. At approximately $\Delta=1.5\hbar\omega$ [i.e.\
$E_{00}+E_{01}=0$, cf.\ Eq.~(\ref{eq:E_nm})] the pair production has
its maximum. A small shift from the simple estimate is due to the
multi-mode character of the problem. The next resonance for the
($n=0,m=0$) level is centered around at $\Delta=2.5\hbar\omega$ [i.e.\
$E_{00}+E_{11}=0$]. Note that around $\Delta=2.5\hbar\omega$, the
production of pairs with ($n=0,m=-1$) and ($n=1,m=2$) is also
resonant. The third resonance, where $E_{00}+E_{21}=0$, is at $\Delta
= 3.5\hbar\omega$.  Here several other resonances are also present.

\begin{figure}[htbp]
  \centering
  \includegraphics{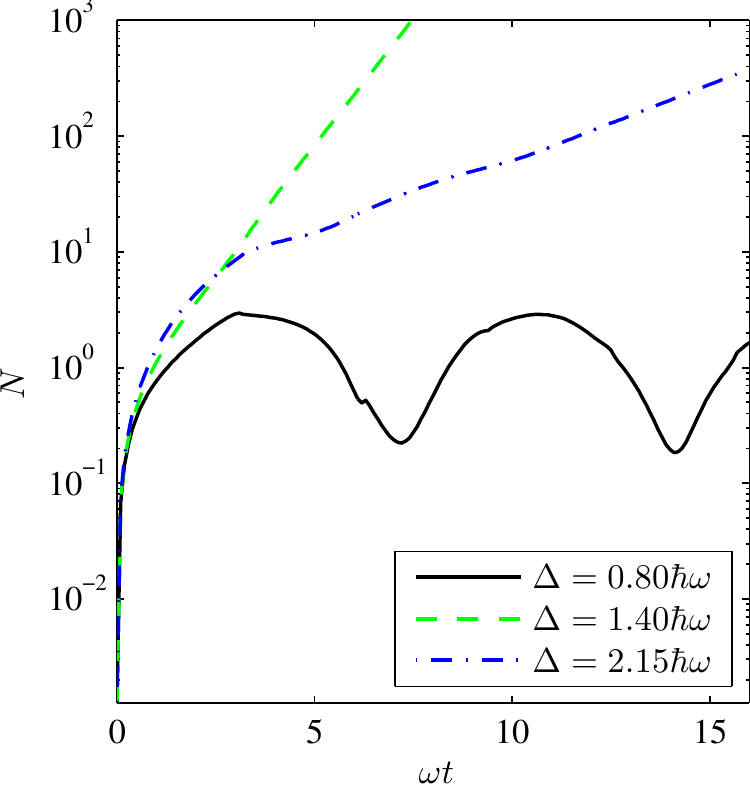}
  \caption{(color online) Total number of atoms as a function of time for
    three values of $\Delta$. The coupling is as in
    Fig.~\ref{fig:lambdas}.}\label{fig:Natoms}
\end{figure}

\begin{figure}[htbp]
  \centering
  \includegraphics{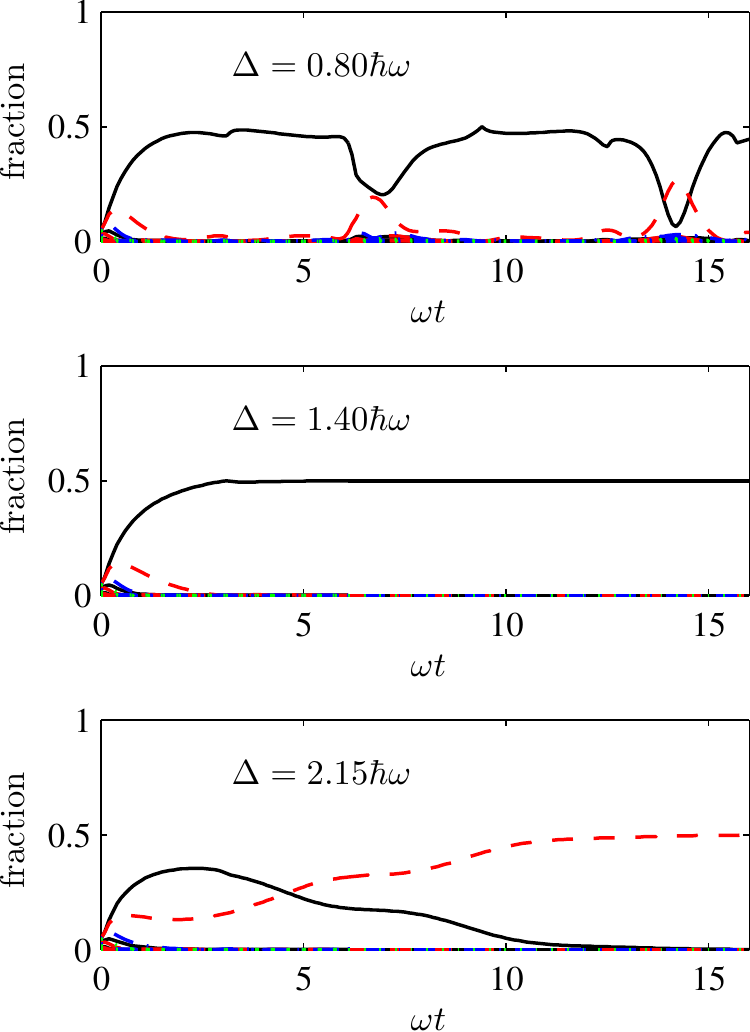}
  \caption{(color online) Condensate fraction, i.e.\ relative weigth
    of the largest eigenvalue of the one-body density operator
    compared to the sum of all eigenvalues (total number of particles.
    Results are plotted for three different values of $\Delta$. As
    discussed in Sec.~\ref{sec:symmetries}, the conservation of
    angular momentum dictates a block-diagonal form of the one-body
    density operator and that pairs of blocks for which $m+m'=M$ will
    have identical spectra. Fully drawn lines correspond to $m=1$ and
    $m'=0$, dashed lines to $m=2$ and $m'=-1$, and dotted lines to
    $m=3$ and $m'=-2$.}\label{fig:cond_frac}
\end{figure}

In Fig.~\ref{fig:Natoms} we plot the total number of atoms and in
Fig.~\ref{fig:cond_frac} the fractional distribution on
single-particle modes as a function of time for three different values
of $\Delta$.
For $\Delta=0.80\hbar\omega$ the dynamics is detuning dominated and
oscillatory.  All modes can be decoupled and the resulting
quasi-particle modes are only slightly squeezed resulting in a very
modest production of atoms.  This almost vanishing population mainly
occupies two modes: one in the $m=0$ manifold and one in the $m'=1$
manifold. These modes constitute the pair that is closest to resonance
and their oscillation period is of the same order of magnitude as the
trap period. The fractional occupation of these modes is close to 50\%
except for times close to the minima in their oscillation (e.g.\
$\omega t\sim 7$ here). At $\Delta=1.40\hbar\omega$, the system can no
longer be decoupled: A single $m=0$, $m'=1$ pair of modes has just
become unstable and real pair production takes place. Therefore this
pair of modes quickly becomes dominant.  Finally, for
$\Delta=2.15\hbar\omega$ the evolution is more complex. From
Fig.~\ref{fig:lambdas} one can tell that there will be real pair
production into a pair of modes with $m=-1$, $m'=2$.  However, a pair
of modes with $m=0,m'=1$ also becomes unstable if $\Delta$ is
increased just a little bit. It turns out that the $m=0,m'=1$ pair
dominates at early times, but later most atoms are produced in the
$m=-1,m'=2$ pair of states.

\begin{figure*}[htbp]
  \centering
  \includegraphics{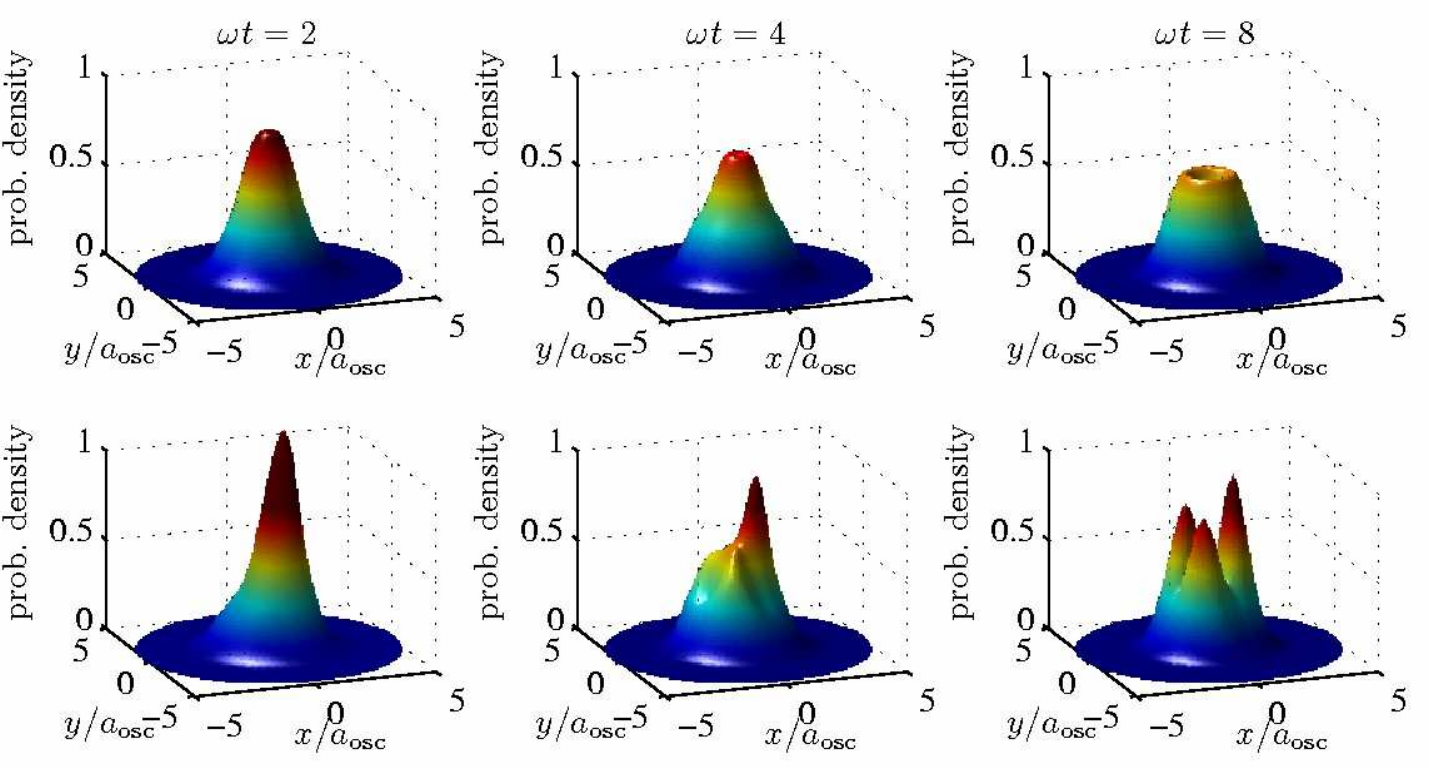}
  \caption{(color online) Detection probability distribution
    (normalized density distribution) before (upper row) and after
    (lower row) detection of a single atom in $(x_d,y_d)=(1,0)\aosc$.
    The detuning $\Delta$ was chosen to be $2.15\hbar\omega$.  The
    three columns show results where the system was allowed to evolve
    for a time $\omega t=2,4$, and $8$, respectively.  Other
    parameters were like in the previous figures.  Before the
    detection of the first atom, the density distribution is
    rotationally symmetric and it is equally likely to make the first
    detection at any azimuthal angle, $\phi$.  According to
    Fig.~\ref{fig:cond_frac}, the one-body density matrix at $\omega
    t=2$ is an almost equal mixture of $m=0$ and $m'=1$ atoms. In each
    of these $m$-manifolds, a single radial mode dominates and
    accounts for 35\% of the total number of atoms. In the lower
    panel, we assume that an atom has been detected with $\phi=0$,
    more precisely at $(x_d,y_d)=(1,0)\aosc$.  The conditional
    probability distribution for the next detection has become
    asymmetric indicating a significantly increased probability to
    detect the next atom close to the first one. Diagonalization shows
    that the one-body density matrix now has a ``condensate fraction''
    of almost 60\%. At $\omega t=4$ a significant number of
    $m=-1,m'=2$ pairs have been created. The density before detection
    has a pronounced dip at the center due to centrifugal effects. The
    conditional density after the detection is rather complex: it is
    in fact not even symmetric around the $y$-axis. Because of the
    larger number of modes involved, the condensate fraction is only
    43\%. Finally, at $\omega t=8$ the $m=-1,m'=2$ pairs dominate.
    After the detection, three almost equally strong peaks are seen,
    signaling the appearence of coherences between two modes with
    angular momentum differing by 3 units. The condensate fraction
    after the detection is now again almost 60\%.}
  \label{fig:conditonal}
\end{figure*}

The above results all concern one-body properties. An illustrative way
to quantify some of the two-body correlations in the system is to
calculate the \emph{conditional} density distribution, i.e., the
density distribution resulting after the detection of a single atom at
some position $(x_d,y_d)$. In Fig.~\ref{fig:conditonal} we plot the
result of such a calculation for $\Delta=2.15\hbar\omega$ and $\omega
t=2,4,6$.
The upper row shows the normalized density distribution, that is, the
probability distribution for the detection of the first atom. In the
lower row, an atom has been detected at $(x_d,y_d)=(1,0)\aosc$ and the
resulting conditional probability distribution for the next detection
is shown. For all three times, this conditional distribution is
dramatically different from the original one. Close to $(x_d,y_d)$,
there is generally a significantly higher probabilty density,
reminiscent of the familiar (thermal state) bunching of bosons.
However, when higher angular momentum states are populated, additional
peaks appear as the detection induces coherences between more $m$
values. From a calculation of the full one-body density operator we
can also find the new ``condensate fraction'' and for all three cases
this number is significantly increased compared to the situation
before the detection (cf. caption of Fig.~\ref{fig:conditonal}). This
is similar to what has been proposed in the case of number state
condensates~\cite{moelmer02:_macros}.

\section{Conclusion}
\label{sec:conclusion}
We have presented a qualitative discussion and a quantitative analysis
of the break-up of a molecular condensate into an atomic system, in
the case where the molecules occupy a vortex state with unit angular
momentum. The rotational symmetry of the problem causes
the fragmentation of the atomic system into (at least) two spatial
components, and our multi-mode analysis quantified the dynamics of
these components. Particle detection at random locations will
generally break the rotational symmetry of the state, and both our
simple two-mode model and our more general analysis show that such
detections will indeed cause the system to approach a single
macroscopically populated quantum state with a preferred angular
dependence.

Several interesting possibilities exist for further studies. A more
technical issue concerns our assumption of non-interacting atoms. As
shown in~\cite{poulsen01:_quant_states_bose_einst} interactions have
non-trivial consequences for the quantum correlations in the system.
Depletion of the molecular condensate~\cite{heinzen00:_super} affects
the quantum statistics of the atomic state, and moreover, the explicit
inclusion of the molecular field will also enable processes where
atoms in different fragments recombine and form molecules with new
angular momenta, respecting always the overall rotational symmetry of
the system, but introducing multi-atom correlations in the system.
Within our simple model with non-interacting atoms and undepleted
molecular systems, it is a natural question to ask, what atomic states
will result from the dissociation of a molecular vortex
\emph{lattice}. Will two fragments be sufficient to describe such
states? We assumed a cylindrical symmetry and referred extensively to
angular momentum conservation in the arguments of this paper, but in
an \emph{interacting} system vortices also exist under non-symmetric
confinement, and our symmetry argument should be adequately modified
into an argument referring to the phase topology.  Presumably one
would still see the formation of a fragmented system with components
with and without vorticity, but the need for interactions among the
atoms for the stability of such a state is an interesting issue.

\begin{acknowledgments}
UVP acknowledges financial support from the Danish Natural Science
Research Council.
\end{acknowledgments}

\appendix*
\section{Cylindrical symmetry}
\label{sec:cyl_sym}

The case of strict cylindrical symmetry can be treated in a very
similar fashion as the one of frozen $z$ dynamics of
Sec.~\ref{sec:frozen_z}. Assume that potential is axially symmetric
and flat along $z$: $\omega_z=0$. The coupling is also assumed to be
$z$ independent, $\chicm(\rho,z)=K(\rho)$. We then use a quantization
box of length $2L$ along $z$ for the atoms. The field operator is
expanded as follows
\begin{multline}
  \label{eq:cyl_expan_oper}
  \hat\Psi(\vr)
  =
  \sum_{n=0}^\infty
  \sum_{m=-\infty}^{\infty}
  \sum_{k=-\infty}^{\infty}
  \hat{a}_{nmk}
  \\
  \times
  \Phi_{nm}(\rho)
  \times
  \sqrt{\frac{1}{2\pi}} e^{im\phi}
  \times
  \sqrt{\frac{1}{2L}}e^{i\pi k z/L}
\end{multline}
where the radial modefunctions where given in Eq.~(\ref{eq:def_phi_nmk}).
The mode creation and annihilation operators satisfy usual commutation
relations
\begin{equation}
  \label{eq:psi_rad_commu}
  \begin{split}
    \Bigl[\hat{a}_{nmk},\hat{a}_{n'm'k'}\Bigr]
    &=
    0
    \\
    \Bigl[\hat{a}_{nmk},\hat{a}^\dagger_{n'm'k'}\Bigr]
    &=
    \delta_{nn'}\delta_{mm'}\delta_{kk'}
    .
  \end{split}
\end{equation}
Under the above assumptions, the Hamiltonian (\ref{eq:H_many_modes})
then simplifies to the form:
\begin{multline}
  \label{eq:H_cyl}
    \hat{H}
    =
    \sum_{nmk}
    E_{nmk} \hat{a}^\dagger_{nmk}\hat{a}_{nmk}
    \\
    +
    \sum_{mk}
    \sum_{nn'}
    \left\{
    K_{nn'm}
    \hat{a}^\dagger_{nmk}\hat{a}^\dagger_{n'(M-m)(-k)}
    +\text{h.c.}
    \right\}
    ,
\end{multline}
where
\begin{equation}
  \label{eq:def_Enmk}
  E_{nmk} = 2n + |m| + 1
  + \frac{1}{2}\left(\frac{\pi}{L}\right)^2k^2
  -\Delta
\end{equation}
and
\begin{equation}
  \label{eq:def_chi_nmk}
  \begin{split}
    &K_{nn'm}
    =
    \\
    &\phantom{K}\int\!  d\!z d\!\rho \rho  \;
    K(\rho) \;
    \frac{1}{2L} \;
    \Phi^*_{nm}(\rho) \Phi_{n'M-m}(\rho)
    .
  \end{split}
\end{equation}

The main difference to the case of a single active $z$ mode treated
above is that excess molecular energy can now be transformed to $z$
kinetic energy of the atoms. Thus the picture of isolated
resonances breaks down, even at low coupling strengths. Many atomic
modes will participate in the dynamics and it will take more
detections to build up a sizable condensate fraction.



\end{document}